\documentstyle[12pt,fleqn,epsfig]{article}
\begin{document}

\input epsf

\begin{center}

{ \bf QUARK-GLUON PLASMA SIGNATURES\\
 IN NUCLEUS--NUCLEUS COLLISIONS AT CERN SPS}

\vspace{0.5cm}

{ \bf Mark I. Gorenstein}

\vspace{0.5cm}

Bogolyubov Institute for Theoretical Physics, Kyiv, Ukraine
\end{center}
\vspace{0.5cm}

\begin{abstract}
\noindent
Two signatures of 
the quark-gluon plasma -- strangeness `enhancement' and $J/\psi$
`suppression' --
in nucleus--nucleus collisions 
at the SPS energies are critically discussed.
\end{abstract}

\vspace{0.5cm}
\noindent
{\bf I. Introduction} 

A discovery of 
the quark-gluon (QGP) 
in nucleus-nucleus (A+A) collisions at CERN SPS 
was recently announced \cite{cern}. 
The present status of this discovery is
still somewhat uncertain
and is right now vigorously debated.
The data on the transverse energy of secondary particles
in central Pb+Pb at 158~A$\cdot$GeV 
and assumed early stage collision geometry lead to
the estimate of the initial energy density 
$\varepsilon_{in}=3\div 4$~GeV/fm$^3$. 
The state of matter at this high energy density
is expected to consist of the deconfined
quarks and gluons.  
Are there evidences of the
transient QGP state in the measured data?

\vspace{0.5cm}
\noindent
{\bf II. Chemical and Thermal Freeze-Out} 

The yields of different hadron
species,
$\pi^+$, $\pi^-$, $K^+$, $K^-$, $p$, $\overline{p}$, ... , $\Omega$,
$\overline{\Omega}$,
have been measured in  
A+A collisions at the SPS energies.
The measured hadron multiplicities are consistent with a simple
picture of the statistical hadronization 
with chemical freeze-out temperature
$T_H= 170\pm 10$~MeV (see Ref.\cite{Br:00}). 
The temperature $T_H$ is close to the 
expected value for the phase transition between hadron matter
and QGP. It suggests a possibility to associate
the chemical freeze-out in A+A collisions
at the SPS energies with QGP hadronization transition.

The data on hadron transverse momentum spectra in Pb+Pb
collisions 
at 158~A$\cdot$GeV
exhibit the exponential behavior
$\exp(-\sqrt{m_i^2+p_{\perp}}/T_i)$ with approximately 
linear dependence of the slope $T_i$ on particle mass $m_i$.
The explanation of these data requires a strong
transverse collective flow, $\langle v_{\perp} \rangle \cong 0.5$,
and low thermal freeze-out temperature $T_f\cong 120$~MeV
(see Ref.\cite{Heinz}).

The data on hadron multiplicities and momentum spectra
give no direct information about the QGP existence. These
hadron observables are formed
after (or during) the QGP hadronization. Any real probe
of the QGP must be present in the early stage of
the reaction (before hadronization) and retain the signature
of the deconfined matter properties throughout the confinement
transition and the subsequent hadron matter evolution. 
Real and virtual photons are emitted at all stages
of the system evolution. They leave the medium without
strong interactions and therefore reflect the matter properties
at the time they were created. Thermal emission of the photons
and lepton pairs can be used to determine the matter temperature
at different stages of A+A collision. The crucial
problem for this probe is however the subtraction of the background
effects: the measured spectra are dominated at high photon
momenta or dilepton mass by hard primary reactions
and at low momenta or masses by hadron decay products.
Thermal photons or dileptons emission from the QGP 
has so far not been identified.

In what follows we discuss to main signals of the QGP:
strangeness `enhancement' and $J/\psi$ `suppression'.   

\vspace{0.5cm}
\noindent
{\bf III. Strangeness `Enhancement'}
 
The idea of strangeness
enhancement as a QGP signal was formulated a long time ago
\cite{raf:86}.
It was based on the estimate that the strangeness equilibration time
in QGP is of the same order ($\approx 10$ fm/c) as
the expected life time of the fireball formated in A+A
collisions.
Thus  in the case of QGP creation the strangeness is expected 
to approach its equilibrium value in QGP.
This equilibrium value is significantly higher
than the strangeness production in 
nucleon--nucleon (N+N) collisions.
Strangeness production in secondary hadronic interactions
was estimated to be negligible small.
Therefore, if QGP is not formed, the strangeness
yields would be expected to be much
lower than those predicted by equilibrium QGP calculations.
Thus at that time a simple and elegant signature of
QGP creation appeared:
{\it a transition to QGP should be signaled 
by an increase of the strangeness
production} to the level of 
QGP equilibrium value.
In an actual study of strangeness production, due to experimental and
theoretical reasons, it is convenient to analyze
strangeness to pion ratio:
\begin{equation}\label{strange}
E_s~=~\frac{\langle \Lambda\rangle ~+~\langle K+\overline{K}\rangle}
{\langle \pi \rangle}~.
\end{equation}
In the QGP picture the ratio can be estimated from the equilibrium
strangeness to entropy ratio using common assumption of isentropic
expansion.

The confrontation of these expectation with the data
was for the first time possible in 1988 when the preliminary
surprising results from S and Si beams at SPS and AGS were presented.
The experiment NA35 reported that in central S+S collisions
at 200 A$\cdot$GeV the strangeness to pion ratio is 2 times higher
than in N+N interactions at the same energy per nucleon.
Even larger enhancement (a factor of about 3)  was measured
by E802 in Si+A collisions at AGS. 
Recent data on central Au+Au collisions at low AGS energies
completed the  picture:
strangeness enhancement is observed at all energies,
it is stronger  at lower energies than at the SPS
energy. This enhancement goes to infinity at the threshold
energy of strange hadron production. 
Thus the AGS  measurements of strangeness enhancement larger
than that at SPS showed clearly that {\it the simple concept
of strangeness enhancement as a signal of QGP is incorrect}.

In fact, for the chemical freeze-out
parameters,
temperature $T$ and baryonic chemical potential $\mu_b$,
found for the SPS energies the strangeness to entropy ratio
is larger in the equilibrium hadron gas (HG) than in the equilibrium QGP.
To estimate the strangeness to entropy
ratio let us consider the quantity
\begin{equation}\label{sentr}   
R_s~\equiv~\frac{N_s+N_{\bar{s}}}{S}~,
\end{equation}
where $N_s$ and $N_{\bar{s}}$ are the numbers of strange quarks and
antiquarks, and S is the total entropy of the system. In the QGP we use
the
ideal gas approximation of
massless {\it u}-, {\it d}-(anti)quarks and gluons, strange
(anti)quarks  with $m_s\cong 150$ MeV. 
For the HG state the values of $N_s$ and
$N_{\bar{s}}$
are calculated as a sum of all $s$ and $\bar{s}$ inside hadrons,
and $S$ is the total HG entropy.
The behavior of $R_s$ (\ref{sentr}) for the HG and QGP
is shown in Fig.~1 as a function of $T$ for $\mu_B=0$.
At temperatures  $T \ge 200$~MeV
one finds in the QGP an almost
constant value of $R_s$ which is smaller than the corresponding
quantity in the HG.
The total entropy as well as    
the total number of strange quarks and antiquarks are expected to
be conserved approximately during the hadronization of QGP.  This
suggests that the value of $R_s$ at the HG chemical freeze-out
should be close to that in the equilibrium QGP and smaller than in the
HG at chemical equilibrium.
Therefore, the strangeness {\it suppression} in the
HG would become a signal for the formation of QGP at the early stage of
A+A collision at the CERN SPS energies.

The statistical model of the early stage of A+A collisions
\cite{early} leads to the following predictions for strangeness
production (see Fig.~2):\\
1. A non-monotonic (or kinky) collision energy dependence
of the strangeness to pion ratio (\ref{strange}). 
A creation of the QGP in the energy region between the
AGS and SPS would change an initial fast increase of this
ratio in equilibrium hadron gas by a {\it decrease} to
the level expected in equilibrium QGP. 
New preliminary data in Pb+Pb at 40 A$\cdot$GeV support this
conclusion 
\cite{40}. \\
2. Very similar strangeness to pion ratio is predicted for
SPS, RHIC and LHC energies as strangeness/entropy ratio
in the QGP is almost independent of temperature (collision energy).

\vspace{0.5cm}
\noindent
{\bf IV. $J/\psi$ `Suppression'}
 
A standard picture of 
$J/\psi$ 
production in hadron and nuclear
collisions assumes a two step process: the creation of
$c\overline{c}$ pair in hard parton collisions at the very early
stage of the reaction and the subsequent formation of
a bound charmonium state. Matsui and Satz proposed \cite{satz:86}
(see also \cite{satz:00} and references therein) to use
$J/\psi$ as a probe for deconfinement in the study of A+A
collisions. They argued that in QGP color screening dissolves
initially created $J/\psi$ mesons into $c$ and $\overline{c}$
quarks which at hadronization form open charm hadrons.
As the initial yield of $J/\psi$ is believed to have
the same A-dependence as the Drell--Yan lepton pairs,
the measurement of a weaker A-dependence of final $J/\psi$ 
yield ($J/\psi$ `suppression')
would signal charmonium absorption and therefore
creation of QGP.
The measured A--dependence of $J/\psi$ production
in p+A is weaker than A$^1$ (approximately A$^{0.9}$).
It was suggested that this $J/\psi$  suppression
is due to absorption in target nucleus. The data
on oxygen and sulphur collisions on nuclei at 200~A$\cdot$GeV
also indicated presence of the considerable suppression.
To improve a fit of the data a new source of $J/\psi$ 
absorption was introduced: the absorption on hadronic secondaries 
(`comovers'). Finally in central Pb+Pb collisions
at 158~A$\cdot$GeV the measured suppression is significantly
stronger than expected in the models including nuclear 
and comover suppressions. This `anomalous' $J/\psi$ 
suppression is now interpreted as an evidence of QGP
creation in Pb+Pb collisions at CERN SPS.

In Ref.~\cite{ga:99} a mechanism of thermal $J/\psi$ production
was suggested. The thermal yield of $J/\psi$ mesons
is given by
\begin{eqnarray}
 \langle J/\psi\rangle ~&=&~ 
\frac{(2j+1)~V}{2\pi^2}~\int_0^{\infty}
\frac{k^2dk}{\exp[(k^2+m_{\psi}^2)^{1/2}/T_H]~-~1} \label{psi} \\
 &\cong & (2j+1)~V~\left(\frac{T_Hm_{\psi}}{2\pi^2}
\right)^{3/2}~\exp\left(-\frac{m_{\psi}}{T_H}\right)~, \nonumber
\end{eqnarray}
where $j=1$ and $m_{\psi}\cong 3.1$~GeV are the spin and mass
of the $J/\psi$ meson and $T_H$ is the hadronization temperature.
The total system volume $V$ is the sum of the proper volume
elements at the hadronization stage. Both $T_H$ and $V$ parameters
are already fitted to the data on hadron yields. Therefore,
Eq.~(\ref{psi}) introduces no additional free parameter.
The ratio 
$\langle J/\psi\rangle /\langle h^-\rangle$
is known experimentally as a function of the mean number of nucleons
$N_P$ participating in
the interaction. The data exhibits approximately
constant value of the ratio for p+p and A+A interactions
including the most recent results on centrality selected Pb+Pb
collisions. The 
$\langle J/\psi\rangle$ 
and $\langle h^-\rangle$ denote the mean multiplicities of
$J/\psi$ mesons and negatively charged hadrons
(more than 90\% are $\pi^-$ mesons), respectively.
A simple assumption of $J/\psi$ thermal production at the 
hadronization stage explains naturally the scaling behavior
of $\langle J/\psi\rangle /\langle h^-\rangle$
ratio and also the absolute number of the produced
$J/\psi$ mesons for $T_H\cong 180$~MeV. 
Within the hard production mechanism
the observed independence of the 
$\langle J/\psi\rangle /\langle h^-\rangle$
ratio of the collision type results as an {\it accidental}
cancelation of several large effects: large initial
$J/\psi$ multiplicity 
in A+A collisions is reduced by a sequence of absorption processes
always (`accidentally') to the scaling (constant) value of the
$\langle J/\psi\rangle /\langle h^-\rangle$
ratio. 

 In the statistical model of Ref.\cite{ga:99} the $J/\psi$ yield is
{\bf independent} of the open charm yield.
Recently  the statistical coalescence model was introduced
for the charmonium production
in Ref.\cite{Br1}.
Similar to the statistical model \cite{ga:99},
the charmonium states are assumed
to be formed at the hadronization stage.
However, they are produced as a coalescence of created earlier
$c$-$\overline{c}$ quarks and therefore
the multiplicities of open and hidden charm
hadrons are {\bf connected} in that model \cite{Br1}.
The numbers of $c$-$\overline{c}$
quarks are restricted to the values expected within the pQCD approach.
It seemed to be larger than the equilibrium HG result.
This requires the introduction of a new parameter in the HG approach
\cite{Br1} --
the charm enhancement factor $\gamma_c$
(it was denoted as $g_c$ in Ref.\cite{Br1}).
This is analogous to the
introduction of strangeness suppression factor $\gamma_s$ \cite{Raf1}
in the HG model, where the total strangeness observed is smaller than its
thermal equilibrium
value. Within this approach the open charm hadron yield is enhanced
by a factor $\gamma_c$ and charmonium yield by a factor $\gamma_c^2$ in
comparison
with the equilibrium HG predictions.

The statistical coalescence model with an exact charm
conservation is formulated in Ref.\cite{Go:00}. The canonical ensemble
suppression effects
are important for the thermal open charm yield even in
the most central Pb+Pb collisions at the SPS energies. These
suppression effects become
crucial when the number of participants $N_p$ decreases.
From the $J/\psi$ multiplicity data in  Pb+Pb collisions at
158~A$\cdot$GeV the
open charm yield is predicted: $N_{c\bar{c}}^{dir} = 0.4\div 0.7$ in
the most central collisions. It is surprisingly close to the estimate
$N_{c\bar{c}}^{eq}
\cong 0.5$ \cite{Br1} for the chemical equilibrium value in the
quark-gluon plasma before hadronization. The model
predicts also the $N_p$-dependence of the open charm
and the yields
of individual open charm states. These predictions
of the statistical coalescence model
(the open charm yield has not been measured in Pb+Pb) can be tested
in the near future (measurements of the open charm are planned
at CERN). Such a comparison will require to specify
more  accurately the  $\langle J/\psi \rangle$ data.

The charm enhancement factor $\gamma_c$ found
from the  $\langle J/\psi \rangle$ data appears to be not much
different from unity.
Therefore, both the statistical model of Ref.\cite{ga:99}
and the statistical coalescence model \cite{Br1,Go:00} lead to
similar results for the $J/\psi$ yield. The predictions of these
two models will differ greatly at RHIC energies:
according to \cite{ga:99} the $J/\psi$ to pion ratio is expected to be
approximately equal to its value at the SPS, but according to
the statistical coalescence model this ratio
should  increase very strongly.

\vspace{0.5cm}
\noindent
{\bf V. Conclusions} 
\begin{itemize}
\item The matter with energy density $\varepsilon_{in}=3\div4$~GeV/fm$^3$
is created at the early stage of central Pb+Pb collisions
at CERN SPS, most probably in the QGP state.

\item The hadronization of the QGP leads to the locally equilibrium
hadron gas state with temperature parameter $T_H= 170\pm 10$~MeV.

\item  
The deconfinement phase transition is expected to occur at the 
collision energies between AGS and SPS
where the strangeness to entropy (pion) ratio in the equilibrium
confined (hadron) matter is higher than in the QGP.
It leads to
non-monotonic (or kinky) dependence
of the strangeness to pion ratio
on collision energy \cite{early} (see Fig.~2).

\item Statistical hadronization of the
QGP is probably an important 
source of $J/\psi$ production \cite{ga:99}.
This fact would open a new look at $J/\psi$ `suppression'
signal of the QGP.
The assumption of $J/\psi$ thermal production at the 
hadronization stage explains naturally the scaling behavior
of $\langle J/\psi\rangle /\langle h^-\rangle$
ratio and also the absolute number of the produced
$J/\psi$ mesons.

\item 
The statistical model of Ref.\cite{ga:99}
and the statistical coalescence model \cite{Br1,Go:00} lead to
similar results for the $J/\psi$ yield at the SPS energies. 
However, the
predictions of these   
two models will differ at RHIC energies:
according to \cite{ga:99} the $J/\psi$ to pion ratio is expected to be
approximately equal to its value at the SPS, but according to
the statistical coalescence model this ratio
should  increase very strongly.

\end{itemize}

\vspace{0.5cm}
{\bf  Acknowledgments.}  I am
thankful to F. Becattini, P. Braun-Munzinger, K.A.~Bugaev,
M. Ga\'zdzicki, 
L. Gerland, W.~Greiner, A.P.~Kostyuk, L.~McLerran,
I.N.~Mishustin, G.C.~Nayak,  K.~Redlich, J.~Stachel
and H.~St\"ocker for useful comments and
discussions.
The financial support of DAAD
 Germany is  acknowledged.
The research described in this publication was made possible in part by
Award \# UP1-2119 of the U.S. Civilian Research \& Development
Foundation for the Independent States of the Former Soviet Union
(CRDF).

\newpage

\begin{figure}[t]\label{fig1}
\mbox{}
\begin{center}
\vfill
\leavevmode
\epsfysize=15cm \epsfbox[100 160 540 760]{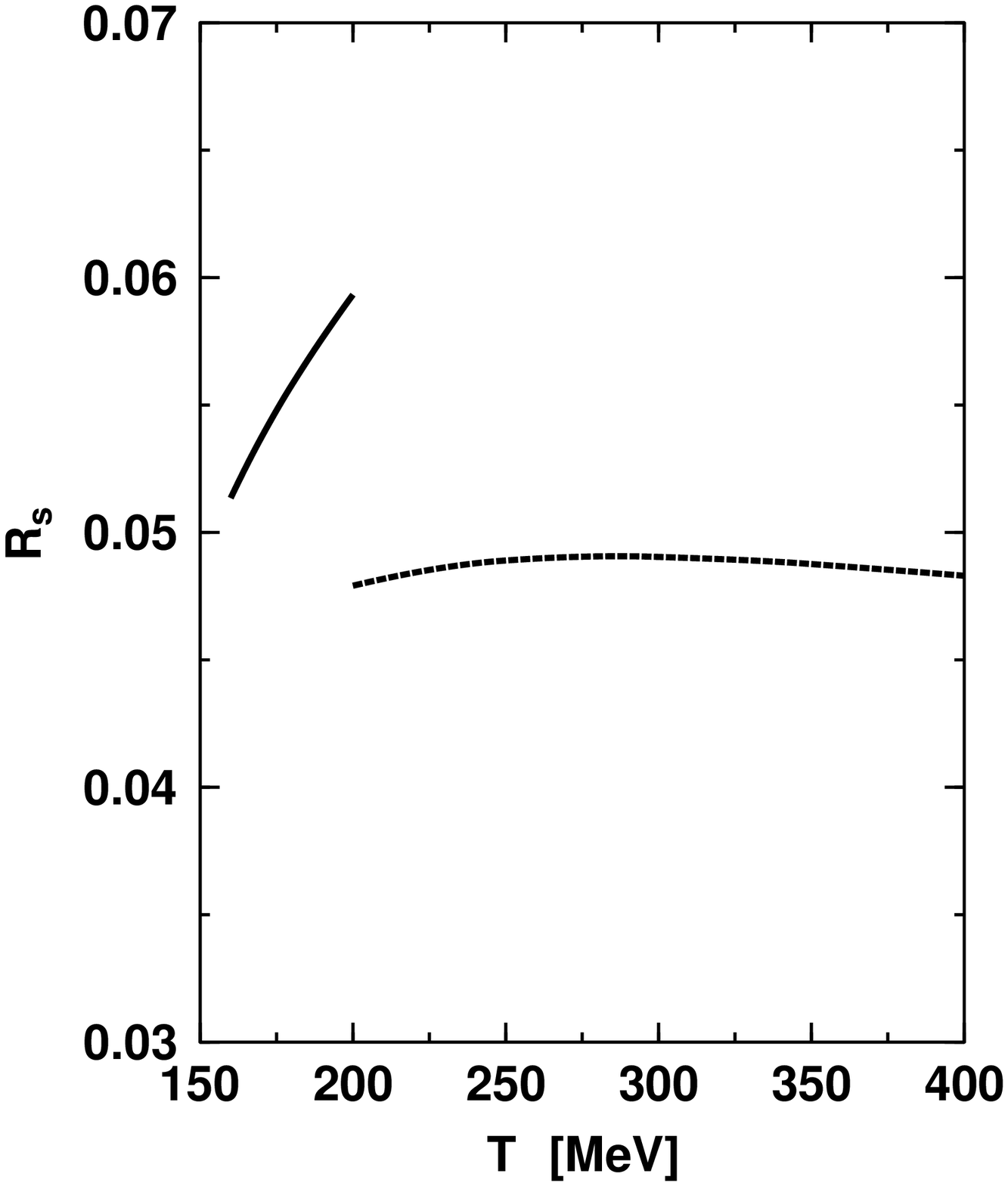}
\vfill
\caption{
$R_s$ (\ref{sentr}) at $\mu_B=0$ for the HG (solid line) and the QGP
(dashed line).
}
\end{center}
\end{figure}

\begin{figure}[t]\label{fig2}
\mbox{}
\begin{center}
\vfill
\leavevmode
\epsfysize=15cm \epsfbox[100 160 540 760]{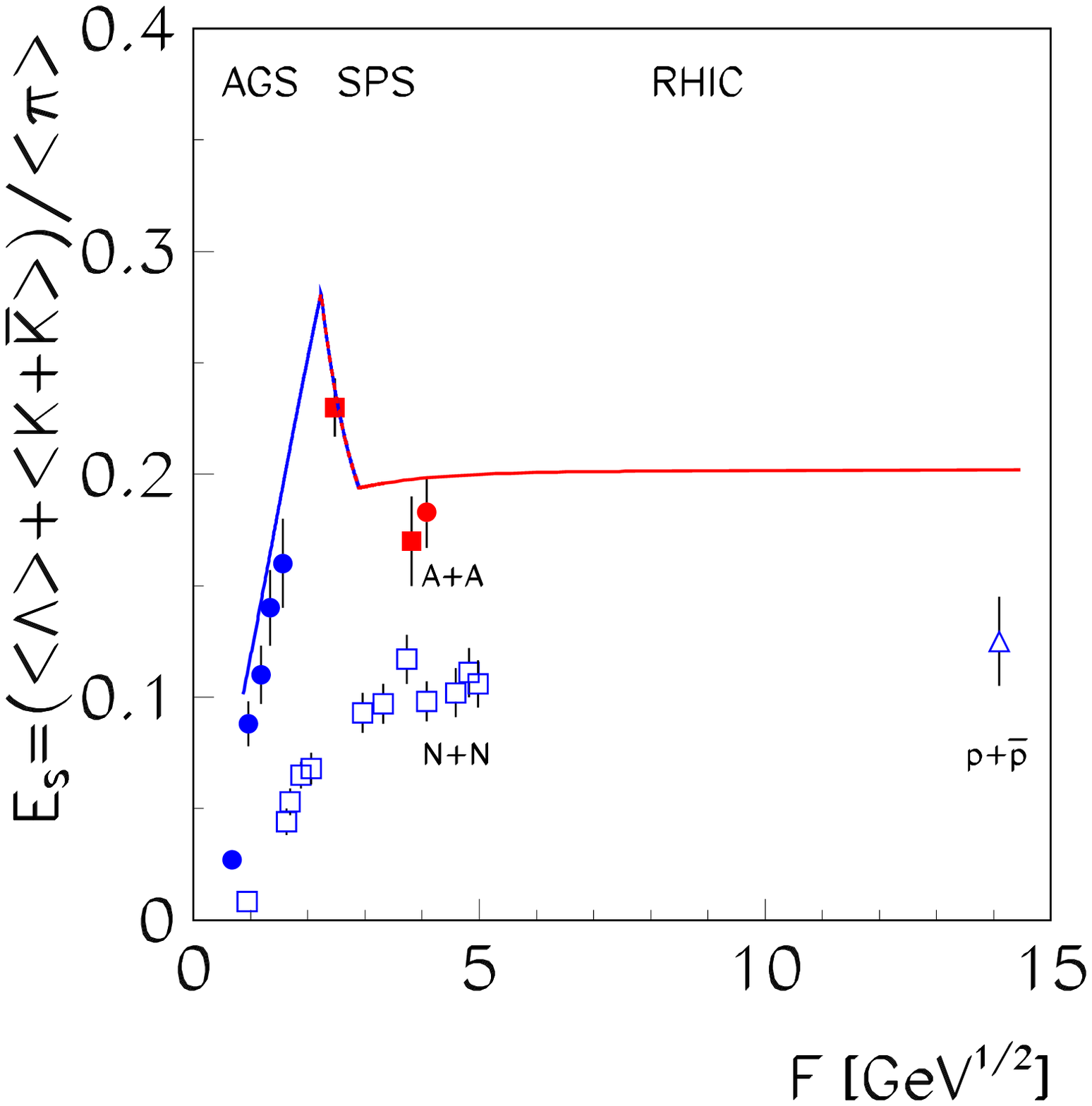}
\vfill
\caption{ Collision energy 
($F\equiv (\sqrt{s}-2m_N)^{3/4}/\sqrt{s}^{1/4}$, $\sqrt{s}$
is nucleon--nucleon c.m. energy)
dependence of strangeness to
pion ratio (\ref{strange}) for central A+A collisions
(closed points), N+N and $p+\overline{p}$ collisions
(open points). The prediction of the statistical model
of Ref.~\cite{early} is shown by solid line.
A transition to the QGP is expected between the
AGS ($F\approx 2$) and the SPS ($F\approx 4$) energies and
leads to the non-monotonic dependence of the strangeness to
pion ratio. 
Preliminary data in Pb+Pb collisions at $E_{lab}=40$~A$\cdot$GeV
\cite{40} presented in the figure seems to support this conclusion.
At high collision energies the ratio saturates at
the value characteristic for equilibrium QGP.}

\end{center}
\end{figure}


\begin{thebibliography}{[99]}

\bibitem{cern}
U. Heinz and M. Jacob,
{\it Evidence for a New State of Matter: 
an Assessment of the Results from the CERN Lead Beam Program},
nucl-th/0002042 (2000);\\
CERN Courier No. 3, page 13, April 2000.


\bibitem{Br:00}
P. Braun-Munzinger, {\it Chemical Equilibration and the
Hadron-QGP Phase Transition}, nucl-ex/0007021 (2000).

\bibitem{Heinz}
U. Heinz, {\it The litle Bang: Searching for quark-gluon plasma in
relativistic heavy-ion collisions}, hep-ph/0009170 (2000).

\bibitem{raf:86} 
P. Koch, B. M\"uller and J. Rafelski,
Phys. Rep. {\bf 142} (1986) 321.

\bibitem{early}
M. Ga\'zdzicki and M. I. Gorenstein,
Acta Phys. Pol. {\bf B30} 2705 (1999).

\bibitem{40}
S.V. Afanasev et al. (NA49 Collab.),
CERN-SPSC-2000-035,
ERN-SPSLC-P-264-ADD-7 (2000).


\bibitem{satz:86}
T. Matsui and H. Satz, Phys. Lett. {\bf B178} (1986) 416.

\bibitem{satz:00}
H. Satz, {\it Colour Deconfinement in Nuclear Collisions},
hep-ph/0007069 (2000).

\bibitem{ga:99}
M. Ga\'zdzicki and M. I. Gorenstein,
Phys. Rev. Lett. {\bf 83} (1999) 4009.

\bibitem{Br1}
P. Braun-Munzinger and J. Stachel,  Phys. Lett. {\bf B490}
(2000) 196.

\bibitem{Raf1}
J. Rafelski, Phys. Lett. {\bf B62} (1991) 333.

\bibitem{Go:00} 
M.I. Gorenstein, A.P.~Kostyuk,
H.~St\"ocker and  W.~Greiner, 
{\it Statistical Coalescence Model
with Exact Charm Conservation}, hep-ph/0010148 (2000)

\end{thebibliography}
\end{document}